\documentclass{aa}
\usepackage{amsmath}
\usepackage[numbers]{natbib}
\usepackage{txfonts}
\usepackage{graphicx}

\begin{document}


\newcommand{\nbb}[0]{{0\nu}\beta\beta}
\newcommand{\dlmm}[0]{d^l_{mm^\prime}(\beta)}
\newcommand{\Cmm}[0]{C_{\mb\msky}(\beta)}
\newcommand{\dconv}[0]{d^l_{m_{\text{b}}m_{\text{sky}}}(\beta)}
\newcommand{\lmax}[0]{l_{\text{max}}}
\newcommand{\mb}[0]{m_{\text{b}}}
\newcommand{\msky}[0]{m_{\text{sky}}}
\newcommand{\mmax}[0]{m_{b\text{max}}}
\newcommand{\Dlmm}[0]{D^l_{mm^\prime}(\alpha,\beta,\gamma)}
\newcommand{\eq}[1]{eq.~(\ref{#1})}
\newcommand{\eqs}[2]{eqs.~(\ref{#1,#2})}
\newcommand{\Eq}[1]{Eq.~(\ref{#1})}
\newcommand{\Eqs}[2]{Eqs.~(\ref{#1},\ref{#2})}
\newcommand{\fig}[1]{fig.~\ref{#1}}
\newcommand{\figs}[2]{figs.~\ref{#1},\ref{#2}}
\newcommand{\Fig}[1]{Fig.~\ref{#1}}
\newcommand{\qbb}[0]{Q_{\beta\beta}}
\newcommand{\tl}[0]{\text{L}}
\newcommand{\tr}[0]{\text{R}}
\newcommand{\nc}[0]{N_{\text{c}}}
\newcommand{\mw}[0]{M_{\text{W}}}
\newcommand{\mz}[0]{M_{\text{Z}}}
\newcommand{\mr}[0]{M_{\text{R}}}
\newcommand{\md}[0]{m_{\text{D}}}
\newcommand{\mn}[0]{m_\nu}
\newcommand{\lh}[0]{\Lambda_{\text{H}}}
\newcommand{\aif}[0]{a^{ff^\prime}_{i;ll^\prime}}
\newcommand{\gs}[0]{\Gamma_{\text{S}}=1}
\newcommand{\gps}[0]{\Gamma_{\text{PS}}=\gamma^5}
\newcommand{\lr}[0]{\Lambda_{\text{R}}}
\newcommand{\wrt}[0]{W_{\text{R}}}
\newcommand{\wl}[0]{W_{\text{L}}}
\newcommand{\ls}[0]{\Lambda_{\text{S}}}
\newcommand{\gf}[0]{G_{\text{F}}}
\newcommand{\mm}[0]{M_{\text{M}}}
\newcommand{\sst}[1]{{\scriptscriptstyle #1}}
\newcommand{\beq}{\begin{equation}}
\newcommand{\eeq}{\end{equation}}
\newcommand{\beqa}{\begin{eqnarray}}
\newcommand{\eeqa}{\end{eqnarray}}
\newcommand{\dida}[1]{/ \!\!\! #1}
\renewcommand{\Im}{\mbox{\sl{Im}}}
\renewcommand{\Re}{\mbox{\sl{Re}}}
\def\simge{\hspace*{0.2em}\raisebox{0.5ex}{$>$}
     \hspace{-0.8em}\raisebox{-0.3em}{$\sim$}\hspace*{0.2em}}
\def\simle{\hspace*{0.2em}\raisebox{0.5ex}{$<$}
     \hspace{-0.8em}\raisebox{-0.3em}{$\sim$}\hspace*{0.2em}}
\def\dn{{d_n}}
\def\de{{d_e}}
\def\datom{{d_{\sst{A}}}}
\def\grhobar{{{\bar g}_\rho}}
\def\gpibar{{{\bar g}_\pi^{(I) \prime}}}
\def\gpibarz{{{\bar g}_\pi^{(0) \prime}}}
\def\gpibaro{{{\bar g}_\pi^{(1) \prime}}}
\def\gpibart{{{\bar g}_\pi^{(2) \prime}}}
\def\mx{{M_X}}
\def\mrho{{m_\rho}}
\def\qpv{{Q_{\sst{W}}}}
\def\lamtv{{\Lambda_{\sst{TVPC}}}}
\def\lamtvs{{\Lambda_{\sst{TVPC}}^2}}
\def\lamtvc{{\Lambda_{\sst{TVPC}}^3}}

\def\bra#1{{\langle#1\vert}}
\def\ket#1{{\vert#1\rangle}}
\def\coeff#1#2{{\scriptstyle{#1\over #2}}}
\def\undertext#1{{$\underline{\hbox{#1}}$}}
\def\hcal#1{{\hbox{\cal #1}}}
\def\sst#1{{\scriptscriptstyle #1}}
\def\eexp#1{{\hbox{e}^{#1}}}
\def\rbra#1{{\langle #1 \vert\!\vert}}
\def\rket#1{{\vert\!\vert #1\rangle}}

\def\lsim{{ <\atop\sim}}
\def\gsim{{ >\atop\sim}}
\def\nubar{{\bar\nu}}
\def\psibar{{\bar\psi}}
\def\Gmu{{G_\mu}}
\def\alr{{A_\sst{LR}}}
\def\wpv{{W^\sst{PV}}}
\def\evec{{\vec e}}
\def\notq{{\not\! q}}
\def\notl{{\not\! \ell}}
\def\notk{{\not\! k}}
\def\notp{{\not\! p}}
\def\notpp{{\not\! p'}}
\def\notder{{\not\! \partial}}
\def\notcder{{\not\!\! D}}
\def\notA{{\not\!\! A}}
\def\notv{{\not\!\! v}}
\def\Jem{{J_\mu^{em}}}
\def\Jana{{J_{\mu 5}^{anapole}}}
\def\nue{{\nu_e}}
\def\mns{{m^2_{\sst{N}}}}
\def\me{{m_e}}
\def\mes{{m^2_e}}
\def\mq{{m_q}}
\def\mqs{{m_q^2}}
\def\mw{{M_{\sst{W}}}}
\def\mz{{M_{\sst{Z}}}}
\def\mzs{{M^2_{\sst{Z}}}}
\def\ubar{{\bar u}}
\def\dbar{{\bar d}}
\def\sbar{{\bar s}}
\def\qbar{{\bar q}}
\def\sstw{{\sin^2\theta_{\sst{W}}}}
\def\gv{{g_{\sst{V}}}}
\def\ga{{g_{\sst{A}}}}
\def\pv{{\vec p}}
\def\pvs{{{\vec p}^{\>2}}}
\def\ppv{{{\vec p}^{\>\prime}}}
\def\ppvs{{{\vec p}^{\>\prime\>2}}}
\def\qv{{\vec q}}
\def\qvs{{{\vec q}^{\>2}}}
\def\xv{{\vec x}}
\def\xpv{{{\vec x}^{\>\prime}}}
\def\yv{{\vec y}}
\def\tauv{{\vec\tau}}
\def\sigv{{\vec\sigma}}

\def\sst#1{{\scriptscriptstyle #1}}
\def\gpnn{{g_{\sst{NN}\pi}}}
\def\grnn{{g_{\sst{NN}\rho}}}
\def\gnnm{{g_{\sst{NNM}}}}
\def\hnnm{{h_{\sst{NNM}}}}
\def\xivz{{\xi_\sst{V}^{(0)}}}
\def\xivt{{\xi_\sst{V}^{(3)}}}
\def\xive{{\xi_\sst{V}^{(8)}}}
\def\xiaz{{\xi_\sst{A}^{(0)}}}
\def\xiat{{\xi_\sst{A}^{(3)}}}
\def\xiae{{\xi_\sst{A}^{(8)}}}
\def\xivtez{{\xi_\sst{V}^{T=0}}}
\def\xivteo{{\xi_\sst{V}^{T=1}}}
\def\xiatez{{\xi_\sst{A}^{T=0}}}
\def\xiateo{{\xi_\sst{A}^{T=1}}}
\def\xiva{{\xi_\sst{V,A}}}
\def\rvz{{R_{\sst{V}}^{(0)}}}
\def\rvt{{R_{\sst{V}}^{(3)}}}
\def\rve{{R_{\sst{V}}^{(8)}}}
\def\raz{{R_{\sst{A}}^{(0)}}}
\def\rat{{R_{\sst{A}}^{(3)}}}
\def\rae{{R_{\sst{A}}^{(8)}}}
\def\rvtez{{R_{\sst{V}}^{T=0}}}
\def\rvteo{{R_{\sst{V}}^{T=1}}}
\def\ratez{{R_{\sst{A}}^{T=0}}}
\def\rateo{{R_{\sst{A}}^{T=1}}}
\def\mro{{m_\rho}}
\def\mks{{m_{\sst{K}}^2}}
\def\mpi{{m_\pi}}
\def\mpis{{m_\pi^2}}
\def\mom{{m_\omega}}
\def\mphi{{m_\phi}}
\def\Qhat{{\hat Q}}
\def\FOS{{F_1^{(s)}}}
\def\FTS{{F_2^{(s)}}}
\def\GAS{{G_{\sst{A}}^{(s)}}}
\def\GES{{G_{\sst{E}}^{(s)}}}
\def\GMS{{G_{\sst{M}}^{(s)}}}
\def\GATEZ{{G_{\sst{A}}^{\sst{T}=0}}}
\def\GATEO{{G_{\sst{A}}^{\sst{T}=1}}}
\def\mdax{{M_{\sst{A}}}}
\def\mustr{{\mu_s}}
\def\rsstr{{r^2_s}}
\def\rhostr{{\rho_s}}
\def\GEG{{G_{\sst{E}}^\gamma}}
\def\GEZ{{G_{\sst{E}}^\sst{Z}}}
\def\GMG{{G_{\sst{M}}^\gamma}}
\def\GMZ{{G_{\sst{M}}^\sst{Z}}}
\def\GEn{{G_{\sst{E}}^n}}
\def\GEp{{G_{\sst{E}}^p}}
\def\GMn{{G_{\sst{M}}^n}}
\def\GMp{{G_{\sst{M}}^p}}
\def\GAp{{G_{\sst{A}}^p}}
\def\GAn{{G_{\sst{A}}^n}}
\def\GA{{G_{\sst{A}}}}
\def\GETEZ{{G_{\sst{E}}^{\sst{T}=0}}}
\def\GETEO{{G_{\sst{E}}^{\sst{T}=1}}}
\def\GMTEZ{{G_{\sst{M}}^{\sst{T}=0}}}
\def\GMTEO{{G_{\sst{M}}^{\sst{T}=1}}}
\def\lamd{{\lambda_{\sst{D}}^\sst{V}}}
\def\lamn{{\lambda_n}}
\def\lams{{\lambda_{\sst{E}}^{(s)}}}
\def\bvz{{\beta_{\sst{V}}^0}}
\def\bvo{{\beta_{\sst{V}}^1}}
\def\Gdip{{G_{\sst{D}}^\sst{V}}}
\def\GdipA{{G_{\sst{D}}^\sst{A}}}
\def\fks{{F_{\sst{K}}^{(s)}}}
\def\FIS{{F_i^{(s)}}}
\def\fpi{{F_\pi}}
\def\fk{{F_{\sst{K}}}}
\def\RAp{{R_{\sst{A}}^p}}
\def\RAn{{R_{\sst{A}}^n}}
\def\RVp{{R_{\sst{V}}^p}}
\def\RVn{{R_{\sst{V}}^n}}
\def\rva{{R_{\sst{V,A}}}}
\def\xbb{{x_B}}
\def\mlq{{M_{\sst{LQ}}}}
\def\mlqs{{M_{\sst{LQ}}^2}}
\def\lscal{{\lambda_{\sst{S}}}}
\def\lvect{{\lambda_{\sst{V}}}}
\def\PR#1{{{\em   Phys. Rev.} {\bf #1} }}
\def\PRC#1{{{\em   Phys. Rev.} {\bf C#1} }}
\def\PRD#1{{{\em   Phys. Rev.} {\bf D#1} }}
\def\PRL#1{{{\em   Phys. Rev. Lett.} {\bf #1} }}
\def\NPA#1{{{\em   Nucl. Phys.} {\bf A#1} }}
\def\NPB#1{{{\em   Nucl. Phys.} {\bf B#1} }}
\def\AoP#1{{{\em   Ann. of Phys.} {\bf #1} }}
\def\PRp#1{{{\em   Phys. Reports} {\bf #1} }}
\def\PLB#1{{{\em   Phys. Lett.} {\bf B#1} }}
\def\ZPA#1{{{\em   Z. f\"ur Phys.} {\bf A#1} }}
\def\ZPC#1{{{\em   Z. f\"ur Phys.} {\bf C#1} }}
\def\etal{{{\em   et al.}}}
\def\delalr{{{delta\alr\over\alr}}}
\def\pbar{{\bar{p}}}
\def\lamchi{{\Lambda_\chi}}
\def\qw0{{Q_{\sst{W}}^0}}
\def\qwp{{Q_{\sst{W}}^P}}
\def\qwn{{Q_{\sst{W}}^N}}
\def\qwe{{Q_{\sst{W}}^e}}
\def\qem{{Q_{\sst{EM}}}}
\def\gae{{g_{\sst{A}}^e}}
\def\gve{{g_{\sst{V}}^e}}
\def\gvf{{g_{\sst{V}}^f}}
\def\gaf{{g_{\sst{A}}^f}}
\def\gvu{{g_{\sst{V}}^u}}
\def\gau{{g_{\sst{A}}^u}}
\def\gvd{{g_{\sst{V}}^d}}
\def\gad{{g_{\sst{A}}^d}}
\def\gvftil{{\tilde g_{\sst{V}}^f}}
\def\gaftil{{\tilde g_{\sst{A}}^f}}
\def\gvetil{{\tilde g_{\sst{V}}^e}}
\def\gaetil{{\tilde g_{\sst{A}}^e}}
\def\gvqtil{{\tilde g_{\sst{V}}^e}}
\def\gaqtil{{\tilde g_{\sst{A}}^e}}
\def\gvutil{{\tilde g_{\sst{V}}^e}}
\def\gautil{{\tilde g_{\sst{A}}^e}}
\def\gvdtil{{\tilde g_{\sst{V}}^e}}
\def\gadtil{{\tilde g_{\sst{A}}^e}}
\def\delp{{\delta_P}}
\def\delzp{{\delta_{00}}}
\def\deld{{\delta_\Delta}}
\def\dele{{\delta_e}}
\def\lnew{{{\cal L}_{\sst{NEW}}}}
\def\osffp{{{\cal O}_{7a}^{ff'}}}
\def\oszg{{{\cal O}_{7c}^{Z\gamma}}}
\def\osgg{{{\cal O}_{7b}^{g\gamma}}}


\def\slash#1{#1\!\!\!{/}}
\def\beq{\begin{eqnarray}}
\def\eeq{\end{eqnarray}}
\def\bea{\begin{eqnarray*}}
\def\eea{\end{eqnarray*}}
\def\NCA{\em Nuovo~Cimento}
\def\IJMP{\em Intl.~J.~Mod.~Phys.}
\def\NP{\em Nucl.~Phys.}
\def\PLB{{\em Phys.~Lett.}~B}
\def\JETPLett{{\em JETP Lett.}}
\def\PRL{\em Phys.~Rev.~Lett.}
\def\MPL{\em Mod.~Phys.~Lett.}
\def\PRD{{\em Phys.~Rev.}~D}
\def\PR{\em Phys.~Rev.}
\def\PRP{\em Phys.~Rep.}
\def\ZPC{{\em Z.~Phys.}~C}
\def\PTP{{\em Prog.~Theor.~Phys.}}
\def\Baryon{{\rm B}}
\def\Lepton{{\rm L}}
\def\sbar{\overline}
\def\stilde{\widetilde}
\def\st{\scriptstyle}
\def\sst{\scriptscriptstyle}
\def\vac{|0\rangle}
\def\argh{{{\rm arg}}}
\def\G{\stilde G}
\def\Wmess{W_{\rm mess}}
\def\NI{\stilde N_1}
\def\antivac{\langle 0|}
\def\infinity{\infty}
\def\mco{\multicolumn}
\def\epp{\epsilon^{\prime}}
\def\psibar{\overline\psi}
\def\nmess{N_5}
\def\chibar{\overline\chi}
\def\lagr{{\cal L}}
\def\drbar{\overline{\rm DR}}
\def\msbar{\overline{\rm MS}}
\def\conj{{{\rm c.c.}}}
\def\Et{{\slashchar{E}_T}}
\def\Etot{{\slashchar{E}}}
\def\mZ{m_Z}
\def\MPlanck{M_{\rm P}}
\def\mW{m_W}
\def\cbeta{c_{\beta}}
\def\sbeta{s_{\beta}}
\def\cW{c_{W}}
\def\sW{s_{W}}
\def\deltaeps{\delta}
\def\sigmabar{\overline\sigma}
\def\epsilonbar{\overline\epsilon}
\def\vep{\varepsilon}
\def\ra{\rightarrow}
\def\half{{1\over 2}}
\def\ko{K^0}
\def\be{\beq}
\def\ee{\eeq}
\def\bea{\begin{eqnarray}}
\def\eea{\end{eqnarray}}
\def\alr{A_{\sst{LR}}}

\def\centeron#1#2{{\setbox0=\hbox{#1}\setbox1=\hbox{#2}\ifdim
\wd1>\wd0\kern.5\wd1\kern-.5\wd0\fi
\copy0\kern-.5\wd0\kern-.5\wd1\copy1\ifdim\wd0>\wd1
\kern.5\wd0\kern-.5\wd1\fi}}
\def\ltap{\;\centeron{\raise.35ex\hbox{$<$}}{\lower.65ex\hbox{$\sim$}}\;}
\def\gtap{\;\centeron{\raise.35ex\hbox{$>$}}{\lower.65ex\hbox{$\sim$}}\;}
\def\gsim{\mathrel{\gtap}}
\def\lsim{\mathrel{\ltap}}
\def\slashchar#1{\setbox0=\hbox{$#1$}           
   \dimen0=\wd0                                 
   \setbox1=\hbox{/} \dimen1=\wd1               
   \ifdim\dimen0>\dimen1                        
      \rlap{\hbox to \dimen0{\hfil/\hfil}}      
      #1                                        
   \else                                        
      \rlap{\hbox to \dimen1{\hfil$#1$\hfil}}   
      /                                         
   \fi}                                        %

\setcounter{tocdepth}{2}







\title{Algorithm for the evaluation of reduced Wigner matrices} 

\author{{G.~Pr{\'e}zeau\inst{1,2}}
\and {M.~Reinecke \inst{3}}}
\institute {Jet Propulsion Laboratory, 4800 Oak Grove Dr., Pasadena, CA 91109, USA \and  California Institute of Technology, 1200 E. California Blvd, Pasadena, CA 91125 \and Max-Planck-Institut f\"ur Astrophysik, Karl-Schwarzschild-Str.~1, 85741 Garching, Germany}



\abstract{Algorithms for the fast and exact computation of Wigner matrices are described and their application to a fast and massively parallel $4\pi$~convolution code between a beam and a sky is also presented.}

\maketitle

\section{Introduction}

Wigner matrices have a ubiquitous presence in science; from the computation of molecular quantum states, through the description of solitons in particle physics and convolution of beam and sky algorithms in astronomy, they are needed to sometimes very high quantum numbers making fast and accurate algorithms that calculate them important.  Other methods have been developed that calculate these matrices exactly but with sub-optimal performance to very high angular momenta~\cite{risbo}, or approximately but very efficiently~\cite{rowe}, but none that calculates them exactly and quickly to almost arbitrarily high angular momentum.  Two such methods are presented in this paper and applied to a convolution algorithm between beam and sky.  The following section gives some basic properties of Wigner matrices, and this is followed by a section describing the algorithm.  The fourth section describes its application to convolution and a summary is presented at the end.

\section{Wigner matrices}

Wigner matrix elements\footnote{For a nice review of Wigner matrices, see \cite{varshalovich}.}, typically denoted by $\Dlmm$, are the eigenfunctions of the Schr\"{o}dinger equation for a symmetric top and form an irreducible basis of the Lie group SU(2), and the rotation group SO(3); the angles $\alpha$, $\beta$ and $\gamma$ are the Euler angles that define the orientation of the top.  As basis functions of SU(2), the $\Dlmm$ satisfy the standard angular momentum relations
\beq\label{llplus1}
\hat{\text{J}}^2\Dlmm &=& l(l+1)\Dlmm \\ \label{mproj}
\hat{\text{J}}_{z}\Dlmm &=& m \Dlmm \\ \label{mprojp}
\hat{\text{J}}_{z^\prime}\Dlmm &=& m^\prime \Dlmm ~,
\eeq
where $l$ labels the irreducible representation of SU(2) and also corresponds to the quantum number representing the total angular momentum of the eigenfunction; $-l\le m,m^\prime\le l$ are the quantum numbers representing the projections of the total angular momentum on two $z$-axes rotated with respect to each other as described below.

The Euler angles are defined as three rotations: a rotation $\gamma$ about the $z$-axis that rotates the $x$ and $y$ axes $\to$ $x^\prime$ and $y^\prime$; this first rotation is followed by a rotation $\beta$ about the new $y^\prime$-axis rotating $x^\prime$ and $z$ axes $\to$ $x^{\prime\prime}$ and $z^\prime$;  the final rotation $\alpha$ is about $z^\prime$.  In the basis we are using as defined by \Eq{mproj} and \Eq{mprojp}, the operators $\hat{\text{J}}_{z}$ and $\hat{\text{J}}_{z^\prime}$ are diagonal and $\Dlmm$ has the form
\beq\label{wignerdef}
\Dlmm = e^{-im\alpha}\dlmm e^{-im^\prime \gamma}
\eeq
where
\beq\label{reducedwignerdef}
\dlmm = \langle lm | \exp\left[ -i\frac{\beta}{\hbar} \hat{\text{J}}_y \right] | lm^\prime \rangle~.
\eeq
$\dlmm$ is called the reduced Wigner matrix element and consists of the overlap of a spherical harmonic with another spherical harmonic that has been rotated by an angle $\beta$ about the y-axis.
The differential equation satisfied by $\dlmm$ is
\beq\label{wigeq}
& &\frac{\text{d}^2 \dlmm}{\text{d}\beta^2} + \text{cot}\beta \frac{\text{d} \dlmm}{\text{d}\beta} +\\ \nonumber
& & ~~~~~~ +\left( \frac{2mm^\prime \cos\beta - m^2 - m^{\prime 2}}{\sin^2\beta} + l(l+1)\right) \dlmm = 0~.
\eeq
From the Schr\"{o}dinger equation in \Eq{wigeq}, it is possible to extract 3-term recursion relations that relate reduced Wigner matrix elements that differ in their quantum numbers.  In principle, it is possible to use such relations to calculate the $\dlmm$.  3-term recursion relations can be unstable, which limits their usefulness unless the potential pitfalls are identified and avoided.  Two examples of these relations are
\beq\label{mrec}
\frac{-m + m^\prime \cos\beta}{\sin\beta} \dlmm = \frac{1}{2}\sqrt{(l+m^\prime)(l-m^\prime +1)}d^l_{mm^\prime-1}(\beta) & & \nonumber
\\
+\frac{1}{2}\sqrt{(l-m^\prime)(l+m^\prime +1)}d^l_{mm^\prime+1}(\beta)~,& & 
\eeq
and
\beq\label{lrec}
\left[ \cos\beta - \frac{mm^\prime}{l(l+1)}\right] \dlmm = \frac{\sqrt{(l^2-m^2)(l^2-{m^\prime}^2)}}{l(2l+1)}d^{l-1}_{mm^\prime}(\beta) & & \nonumber
\\
+\frac{\sqrt{[(l+1)^2-m^2][(l+1)^2-{m^\prime}^2]}}{(l+1)(2l+1)}d^{l+1}_{mm^\prime}(\beta)~.& & 
\eeq

Generally, 3-term recursion relations will have two linearly independent solutions, $f_n$ and $g_n$~\cite{recipes};  these solutions can be oscillatory or exponentially decreasing or increasing.  In the non-oscillatory case,  $f_n$ is the {\it minimal} solution if
\beq
\frac{f_n}{g_n} \to 0 ~~~\text{as}~~~ n \to \infinity~,
\eeq
while $g_n$ is the {\it dominant} solution.  For solutions to the Schr\"{o}dinger equation, exponentially increasing/decreasing solutions appear only in the region where a particle can not classically exist because of energy conservation, but where a wave function can be non-zero in quantum mechanics.  In the case of a rigid rotor~\cite{edmonds}, the kinetic energy of a spherically symmetric rotor is:
\beq\label{classic}
2IT = p_{\beta}^2 +  \frac{1}{\sin^2\beta} (p_{\gamma}^2+p_{\alpha}^2 -2p_{\alpha}p_{\gamma}\cos\beta )
\eeq
In classical mechanics, $p_{\beta}^2>0$.  In quantum mechanics, the quantization of \Eq{classic} means substituting $p_{\alpha}\to -i\partial/\partial\alpha$, $p_{\beta}\to -i\partial/\partial\beta$ and $p_{\gamma}\to -i\partial/\partial\gamma$.  These substitutions combined with an eigenfunction of the form (\ref{wignerdef}) and the additional substitution $2IT\to l(l+1)\Dlmm$ inferred from \Eq{llplus1} yields \Eq{wigeq} .  Since $p_{\beta}^2$ corresponds to the first two terms of \Eq{wigeq}, one concludes that classically we would have
\beq\label{classiccond}
 l(l+1) + \frac{2mm^\prime \cos\beta - m^2 - m^{\prime 2}}{\sin^2\beta} \ge 0 ~ .
\eeq
Wherever \Eq{classiccond} is not satisfied, the solutions will be exponentially suppressed or divergent.  When solving the Schr\"{o}dinger equation for the physical solutions, the divergent solutions are simply put to zero.  When using the 3-term recursion relations, the divergent solution can be 'sniffed' out because of round-off errors and the recursions quickly fail.  One special case where this cannot happen is when $m,m^\prime=0$ where \Eq{classiccond} is always satisfied since $l\ge 0$.  In that case, \Eq{lrec} is stable and can be used to calculate $d^l_{00}(\beta)$ to very high $l$ extremely accurately.

For the cases where $m,m^\prime\ne 0$, we can still use 3-term recursion relations provided we do so in the right direction in the quantum number being varied.  For example, looking at \Eq{mrec}, one can either calculate each $\dlmm$ for increasing $m^\prime$ or decreasing $m^\prime$.  In one direction, the divergent solution will be growing while in the other it will be shrinking.  To determine the direction in which \Eq{mrec} is stable, one need only consider \Eq{classiccond}.  Assume you are interested in evaluating all the reduced Wigner matrix elements $d^l_{0m^\prime}(\beta)$ for $0\le m^\prime \le l$ using \Eq{mrec}; you can choose to begin your recurrence with $d^l_{00}(\beta)$ and increasing $m^\prime$ or begin from $d^l_{0l}(\beta)$ and decreasing $m^\prime$.  To use \Eq{mrec} in a stable manner, you need start from $d^l_{0l}(\beta)$ and decrease $m^\prime$.  Putting $m=0$ in \Eq{classiccond} yields the new condition
\beq\label{classiccondm0}
 l(l+1) - \frac{ m^{\prime 2}}{\sin^2\beta} \ge 0 ~ ,
\eeq
where it is seen that as $m^\prime$ increases from 0 (taking for example $\beta=\pi/4$), we approach the non-classical and violate \Eq{classiccondm0} at $m^\prime \ge \sin\beta\sqrt{l(l+1)}$; increasing $m^\prime$ further means sampling the divergent dominant solution of the Schr\"{o}dinger equation from which \Eq{mrec} is derived.  It is then clear that the stable direction to use \Eq{mrec} is for decreasing $|m^\prime|$.  From \Eq{classiccond}, it is seen quite generally that the recursion relations (\ref{mrec}) and (\ref{lrec}) will be stable provided they are used in the direction of decreasing $| m^\prime |$ and increasing $l$ respectively.

\begin{figure}
\resizebox{8.7 cm}{!}{\includegraphics*{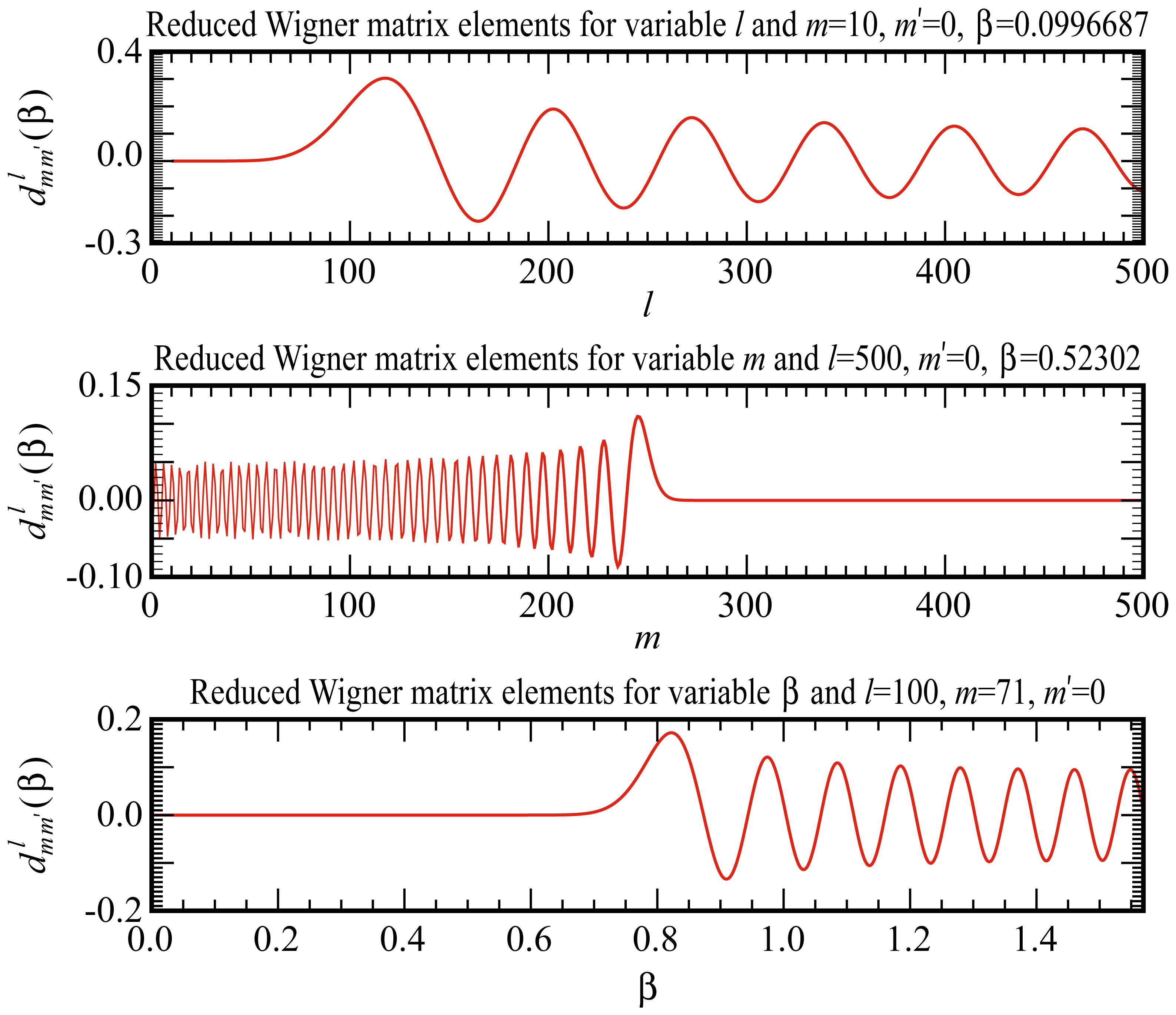}}
\caption{The top plot shows the variation of $d^{l}_{010}(0.0996687)$ for $10\le l\le 500$; the middle plot shows the variation of $d^{500}_{0m^\prime}(0.52331)$ for $0\le m^\prime\le 500$; the third plot shows the variation of $d^{100}_{071}(\beta)$ for $0\le\beta\le \pi/2$.}\label{dlmmplots}
\end{figure}

In addition to Eqs~(\ref{mrec}) and (\ref{lrec}), a third recursion relation in $\beta$ can be derived by discretizing the derivatives in \Eq{wigeq} with the relations
\beq\label{firstder}
f^\prime(x) &\cong& \frac{f(x+\epsilon) - f(x-\epsilon)}{2\epsilon} + O(\epsilon^2 f^{\prime\prime\prime}) \\
f^{\prime\prime}(x) &\cong& \frac{f(x+\epsilon) + f(x-\epsilon) - 2f(x)}{\epsilon^2} + O(\epsilon f^{\prime\prime\prime})~.
\eeq
Substituting into \Eq{wigeq} yields
\beq\label{betarec}
& &\left[\epsilon^2\left( \frac{2mm^\prime \cos\beta - m^2 - m^{\prime 2}}{\sin^2\beta} + l(l+1)\right) - 2 \right] d^l_{mm^\prime}(\beta) \cong \nonumber \\
& & ~~ \left( \frac{\epsilon\text{cot}\beta}{2} - 1 \right) d^l_{mm^\prime}(\beta - \epsilon) - \left( \frac{\epsilon\text{cot}\beta}{2} + 1 \right) d^l_{mm^\prime}(\beta + \epsilon) +  \\ \nonumber
& & ~~~~~~~~~~~~~~~~~+ O(\epsilon^3 d^{l^{\prime\prime\prime}}_{mm^\prime}(\beta))~.
\eeq
From \Eq{classiccond}, it is seen that this recursion relation should be used for increasing $\beta$ if $0<\beta<\pi/2$ and decreasing $\beta$ if $\pi/2<\beta<\pi$.  Examples of these conclusions are given in \Fig{dlmmplots}.  The top plot shows the change in the behavior of $\dlmm$ with increasing $l$ as one moves from the non-classical to classical regions; in that case, the angle $\beta$ was chosen so that \Eq{classiccond} is satisfied only when $l\ge 100$.  The middle plot shows the variation of $\dlmm$ with $m^\prime$.  With $\beta=0.52331$, $l=1000$ and $m=0$, the transition from non-classical to classical regimes occurs at $m^\prime = 500$.  The last plot shows the variation of $\dlmm$ with $\beta$; the value of $m^\prime=71$ was chosen so that the transition from non-classical to classical occurs at $\beta=\pi/4$.  A noticeable feature of all three plots is the tallest peak is always the first peak after the transition to the classical region.  This is qualitatively understandable from \Eq{reducedwignerdef} where the reduced Wigner matrix is seen to characterize the overlap between two spin states after a rotation.  In the classical limit of large $l$, the angle $\omega$ of the spin direction of a quantum object with the $z$ axis is given by
\beq\label{classicalAng}
| lm \rangle ~~ \text{:} ~~ \cos\omega \approx {m\over\sqrt{l(l+1)}}~.
\eeq
One might expect that the overlap would be greatest when the 'classical' spins are aligned after the rotation about the $y$-axis.  From Eqs.~(\ref{classiccond}) and (\ref{classicalAng}), we can show that this is the case when
\beq\label{betaclassic}
\beta=\text{acos}\left(\frac{m^\prime}{\sqrt{l(l+1)}}\right)-\text{acos}\left(\frac{m}{\sqrt{l(l+1)}}\right)~.
\eeq
In our example, $m=0$ and \Eq{betaclassic} reads $\sin(\beta)=m^\prime/[l(l+1)]$ and the overlap is greatest at the transition point.

\section{Algorithm}

The evaluation of the $\dlmm$ from Eqs.~(\ref{mrec}) and (\ref{lrec}) requires starting values for the recursions.  For large $l$ however, those values are often vanishingly small and cannot
be represented by any of the IEEE 754 floating-point data formats which are used on practically all current computer hardware. Starting the recursions with 0 and 1 does not help because there will come a point where the $\dlmm$ become too big to be represented numerically.  Two solutions to this problem are presented here.

\subsection{$\dlmm$ ratios}

From \Eq{wigeq} and the plots of \Fig{dlmmplots}, it is clear that the $\dlmm$ vary smoothly with varying $l$, $m$, and $\beta$.  As a result, a recursion relation of ratios should always be finite in the non-classical region where the $\dlmm$ are not oscillatory, and one only has to worry about singularities in the ratios in the classical/oscillatory region where the denominators could vanish if evaluated at a zero of the $\dlmm$.  In this ratio-based method, Eqs.~(\ref{mrec}) and (\ref{lrec}) can be rewritten
\beq\label{ratiom}
& &\frac{d^l_{mm^\prime}}{d^l_{mm^\prime-1}} =    \\ \nonumber
& &  ~~\frac{  \sqrt{ (l+m^\prime)(l-m^\prime+1) }     }{ m\,\text{cosec}\beta - m^\prime\text{cotan}\beta -   \sqrt{ (l-m^\prime)(l+m^\prime+1)}  \frac{  d^l_{mm^\prime+1}  }{   d^l_{mm^\prime}   }   } \\ \label{ratiol}
& &\frac{d^l_{mm^\prime}}{d^{l+1}_{mm^\prime}}  =    \\ \nonumber
& &   \frac{  (l\!+\!1) \sqrt{ (l^2\!-\!{m^\prime}^2)(l^2\!-\!m^2) }     }{ (2l\!+\!1)( mm^\prime\! -\! \cos\!\beta ) -   l \sqrt{ [(l\!+\!1)^2\!-\!{m^\prime}^2][(l\!+\!1)^2\!-\!m^2] }  \frac{   d^{l-1}_{mm^\prime}   }{   d^l_{mm^\prime}   }    }  ~.
\eeq
Using
\beq
\frac{d^l_{ml}}{d^l_{ml-1}} &=& \frac{  \sqrt{l/2}\sin\beta  }{  l\cos\beta - m}~~\text{and} \\
\frac{d^l_{ml}}{d^{l+1}_{ml}} &=& \sqrt{  \frac{   (l+m+1)(l-m+1)   }{   2l+1  }  }  [ (l+1)\cos\beta - m ]^{-1}~,
\eeq
the ratios can be calculated recursively down to $m^\prime=0$ if using \Eq{ratiom} or up to a $l=\lmax$ if using \Eq{ratiol}.  For example, in the case where all the $d^l_{0m^\prime}$ for $l\ge m^\prime \ge 0$ are required, one would start with \Eq{ratiom} to calculate:
\beq\label{ratiosm0}
\frac{d^l_{0l}(\beta)}{d^l_{0l-1}(\beta)},~\frac{d^l_{0l-1}(\beta)}{d^l_{0l-2}(\beta)},~\dots,~\frac{d^l_{02}(\beta)}{d^l_{01}(\beta)},~\frac{d^l_{01}(\beta)}{d^l_{00}(\beta)}~.
\eeq
To then calculate the $d^l_{0m^\prime}$, one would need to know $d^l_{00}$.  Fortunately, the $d^l_{00}$ are easy to calculate because their recursion relation do not contain exponential solutions as remarked under \Eq{classiccond}:
\beq\label{dl00}
\cos\beta d^l_{00}(\beta)=\frac{l}{2l+1}d^{l-1}_{00} + \frac{l+1}{2l+1}d^{l+1}_{00}
\eeq
Once $d^l_{00}(\beta)$ has been calculated, $d^l_{01}$ can be calculated from the ratios; in order, each $d^l_{0m}(\beta)$ can be calculated by multiplying adjacent ratios until $d^l_{0l}(\beta)$ has been evaluated.  In the case where $d^l_{1m^\prime}$ for $l\ge m^\prime \ge 0$ are also needed, the set of ratios
\beq\label{ratiosm1}
\frac{d^l_{1l}(\beta)}{d^l_{1l-1}(\beta)},~\frac{d^l_{1l-1}(\beta)}{d^l_{1l-2}(\beta)},~\dots,~\frac{d^l_{12}(\beta)}{d^l_{11}(\beta)},~\frac{d^l_{11}(\beta)}{d^l_{10}(\beta)}~.
\eeq
are next computed.  To then calculate the $d^l_{1m^\prime}$, one needs to know $d^l_{10}$.  Fortunately, $d^l_{01}=-d^l_{10}$ has previously been calculated and all the $d^l_{1m^\prime}$ can  be obtained up to $d^l_{1l}$.  In this fashion, all the $\dlmm$ can be calculated up to a desired $m=m_{\text{max}}$.

The special and extremely rare case where a ratio $d^l_{mm^\prime+1}/d^l_{mm^\prime}$ is infinite can be handled by using \Eq{mrec} where $\dlmm$ is set to zero and substituting an infinite ratio and a null ratio for a single finite ratio:
\beq
 \frac{d^l_{mm^\prime+1}}{d^l_{mm^\prime}},~\frac{d^l_{mm^\prime}}{d^l_{mm^\prime-1}}  \to  \frac{d^l_{mm^\prime+1}}{d^l_{mm^\prime-1}} = -\frac{\sqrt{(l+m^\prime)(l-m^\prime+1)}}{\sqrt{(l-m^\prime)(l+m^\prime-1)}}
\eeq
Note that in contrast to the method described in \cite{risbo}, the column of matrix elements $d^l_{ml}(\beta)\to d^l_{m0}(\beta)$ can be evaluated without having to calculate every single $d^{l^\prime}_{mm^\prime}(\beta)$ for $l^\prime < l$.
The same tricks can be applied to the recursion relation in $l$ for the evaluation of $d^l_{lm^\prime}(\beta)\to d^{\lmax}_{lm^\prime}(\beta)$.  First calculate the column of elements $d^{\lmax}_{l\lmax}(\beta)\to d^{\lmax}_{l0}(\beta)$ and then calculate the ratios
\beq\label{ratiosl0}
\frac{d^l_{lm^\prime}(\beta)}{d^{l+1}_{lm^\prime}(\beta)},~\frac{d^{l+1}_{lm^\prime}(\beta)}{d^{l+2}_{lm^\prime}(\beta)},~\dots,~\frac{d^{\lmax-2}_{lm^\prime}(\beta)}{d^{\lmax-1}_{lm^\prime}(\beta)},~\frac{d^{\lmax-1}_{lm^\prime}(\beta)}{d^{\lmax}_{lm^\prime}(\beta)}~.
\eeq
Knowing $d^{\lmax}_{lm^\prime}(\beta)$ allows us to evaluate $d^{\lmax-1}_{lm^\prime}(\beta)\to d^l_{lm^\prime}(\beta)$.

For a particular $\beta$, it is sufficient to compute the elements of $\dlmm$ for $0 \le m \le l$ and $-l \le m^\prime \le l$ to know the entire matrix $d^l(\beta)$.  To fill out the rest of the matrix, the symmetry relations in appendix~\ref{appa} can be used.

\subsection {Wigner matrix elements by $l$-recursion}\label{wigl}

Another way to deal with the underflow problem is to start from \Eq{lrec} with the following initialization values
\begin{align}
 && &d^l_{l,m}(\beta)& &=& &A \, \left( \cos\frac{\beta}{2}\right)^{l+m} \, \left(-\sin\frac{\beta}{2}\right)^{l-m}& &&\label{ini1}\\
 && &d^l_{-l,m}(\beta)& &=& &A \, \left( \cos\frac{\beta}{2}\right)^{l-m} \, \left(\sin\frac{\beta}{2}\right)^{l+m}& &&\\
 && &d^l_{-l,m}(\beta)& &=& &A \, \left( \cos\frac{\beta}{2}\right)^{l+m} \, \left(\sin\frac{\beta}{2}\right)^{l-m}& &&\\
 && &d^l_{-l,m}(\beta)& &=& &A \, \left( \cos\frac{\beta}{2}\right)^{l-m} \, \left(-\sin\frac{\beta}{2}\right)^{l+m}\text{,}&  &&\label{ini2}
\end{align}
where $A=\sqrt{(2l)!/(l+m)!(l-m)!}~$.  As far as equations (\ref{ini1}) to (\ref{ini2}) are concerned, the underflow problem
can be avoided by simply calculating the logarithm of the absolute value
of the matrix element and storing its sign separately.
Equation (\ref{ini1}), e.g., then transforms to
\begin{eqnarray}
  \ln \left|d^l_{l,m}(\beta)\right|&=&0.5\left(\ln((2l)!)-\ln((l+m)!)-\ln((l-m)!\right)\nonumber \\
   &+& (l+m)\ln\left|\cos(\beta/2)\right| + (l-m)\ln\left|\sin(\beta/2)\right| \label{logstart}
\end{eqnarray}

In cases where one of the last two terms is $-\infty$, the recursion in $l$ can be stopped
immediately, since all subsequent values will be zero.

The logarithms of the faculties are easily precomputed, so that the seed
value for the recursion can be obtained in $\mathcal O(1)$ operations.

Since the result of eq.\ (\ref{logstart}) is in some circumstances much smaller
than the individual terms on the right-hand side, cancellation errors may
reduce the number of significant digits of the result.
In order to have the highest accuracy that can be achieved without sacrificing
too much performance, the computation of the seed value is carried
out with extended IEEE precision (corresponding to the C++ data type
{\tt long double}).

The recursion relation (\ref{lrec}) itself unfortunately cannot be computed
conveniently in logarithms; therefore a way must be found to represent floating
point values with an extreme dynamic range, which does not incur a high
performance penalty.

This was implemented by representing a floating-point number $v$ using an
IEEE double precision value $d$ and an integer scale $n$, such that
\begin{equation}
v=d\cdot S^n\text{,}
\end{equation}
where either $d=0$, or $S^{-1}\le|d|\le S$ and $S$ (the ``scale factor'') is a
positive constant that can be represented as a double-precision IEEE value.
Using this prescription, $v$ does not have a unique representation as a
$(d,n)$-pair, but this is not a problem.

Similar techniques have been in use since at least three decades in numerical
algorithms; for a recent example see the spherical
harmonic transform routines of the HEALPix package.

It is advantageous to choose a scale factor which is an integer power of 2,
because multiplying or dividing by such a factor only affects the exponent of a
floating-point value stored in binary format, and is therefore exact
(ignoring possible under- or overflows).
In order to avoid frequent re-scaling of $d$, the scale factor should also be
rather large; the value adopted for our implementation  is $2^{90}$.

Using this representation for the $d^l_{mm'}(\beta)$, the recursion is
performed until either $\lmax$ is reached, or the matrix element has
become large enough to be safely represented by a normal double-precision
variable (the threshold value used in the code is $2^{-900}$).
In the latter case the remaining computations up to $\lmax$
are done with standard floating-point arithmetic, which is signi\-fi\-cant\-ly faster.

\section{Convolution}

One area where fast and efficient techniques of computing $\dlmm$ are particularly valuable is in $4\pi$ convolution~\cite{Wandelt:2001gp}.  For the convolution of two fields $b(\Omega)$ and $s(\Omega)$ defined on a sphere, the following integral must be calculated:
\beq
c = \int\text{d}\!\Omega b^*\!(\Omega)s(\Omega)
\eeq
In the physical application where $b(\Omega)$ is a beam from a horn located on a slowly rotating space telescope that scans the sky (denoted $s(\Omega)$) as it orbits the sun (WMAP or Planck missions, e.g.), a large number of convolutions must be performed to account for every possible orientation $(\alpha, \beta, \gamma)$ of the satellite 
\beq\label{conv}
c(\Omega^\prime) = \int\text{d}\!\Omega [\text{R}(\Omega^\prime)b(\Omega) ]^*s(\Omega)
\eeq
where $\text{R}(\Omega^\prime)$ is a rotation matrix that rotates the beam to a particular orientation of the satellite and is defined
\beq\label{rot}
\text{R}(\alpha, \beta, \gamma) \text{Y}_{lm}(\theta,\phi) = \sum_{m^\prime=-l}^{l} D^l_{m^\prime m}(\alpha, \beta, \gamma) \text{Y}_{lm^\prime}(\theta,\phi) ~,
\eeq
and where the $\text{Y}_{lm}(\theta,\phi)$ are spherical harmonics.  The $ \text{Y}_{lm}(\theta,\phi)$ are related to the $\Dlmm$ through the relation
\beq
 \text{Y}_{lm}(\theta,\phi) = (-1)^m \sqrt{ \frac{  2l+1  }{ 4\pi } } D^l_{0m}(0,\theta,\phi)
\eeq
and can therefore be calculated using the methods described above.

For beams with significant side-lobes stemming from reflections of light far away from the line of sight as is the case for both the WMAP and Planck missions, the beams can cover a significant portion of the sky and full-sky convolutions are necessary; as shown in ref~\cite{Wandelt:2001gp}, such full-sky convolutions are much faster when performed in harmonic space instead of pixel space.  We now describe a very fast and massively parallel method to perform full-sky convolutions in harmonic space.

\subsection{\tt CONVIQT}

{\tt conviqt} (CONvolution VIa the Quantum Top equation) is a fast 4$\pi$ convolution algorithm that relies on fast computational methods for reduced Wigner matrix elements.  Starting from Eqs.~(\ref{conv}) and (\ref{rot}), the beam and sky fields can be expanded on the spherical harmonic basis to yield
\beq\label{conviqt}
c (\alpha, \beta, \gamma) &=&  \sum_{\mb=-\mmax }^{\mmax}\sum_{\msky=-\lmax }^{\lmax}   e^{-i\mb\alpha}e^{-i\msky \gamma} C_{\mb\msky}(\beta) ~, \\ \label{convdef}
C_{\mb\msky}(\beta) &\equiv& \sum_{l=0}^{\lmax} {b^*_{l\mb}} \dconv s_{l\msky}~,
\eeq
where $b_{l\mb}$ and $s_{l\msky}$ are the spherical components of the beam and sky fields.\footnote{For ease of reading, only the scalar case is described since the generalization to polarized maps and beams is easily accomplished by evaluating \Eq{convdef} for the additional pairs $(b_{l\mb}^G,~s_{l\msky}^G)$ and $(b_{l\mb}^C,~s_{l\msky}^C)$.}  Typically, the $b_{l\mb}$ are negligible for some $\mmax<\mb\ll\lmax$.  Noting that the number of $\beta$ angles needed for the convolution scales as $\lmax$, the evaluation of \Eq{conviqt} scales as $O(\lmax^2\mmax\log(\lmax))$ after the use of the Fast Fourier Transform algorithm to perform the summations.  The numerically expensive part of \Eq{conviqt} is the computation of \Eq{convdef} which scales as $O(\lmax^3\mmax)$.  Two separate computations of \Eq{convdef} scale as $O(\lmax^3\mmax)$: the computation of the $\dconv$, and the evaluation of the sum over $0\le l \le \lmax$ for every single $m$, $\msky$, and $\beta$. The fast methods described in the previous section are used to compute the $\dconv$.  To evaluate the sum for each $\beta$, a massively parallel MPI-based approach is used since the $C_{\mb\msky}(\beta)$ are uncorrelated between the different $\beta$ and can be computed by different tasks.  Additional acceleration techniques for both the computation of the $\dconv$ and for the evaluation of the sums over $l$ are described in the following sub-section.

\subsection{Acceleration techniques for the $\dconv$}

In simulations of the measurement of cosmic microwave background, convolutions appear repeatedly especially if Monte Carlos are required.
Since the generation of Wigner matrix elements is typically the most computationally
intensive part of the convolution algorithm, a large effort was made to
increase its efficiency.  This has two aspects: first to compute the matrix
elements as quickly as possible, but also to decide (if possible) which matrix
elements are too small to contribute measurably to the result and skip their
generation altogether.

\subsubsection{Skipping unneeded calculations}

When performing convolutions, especially within the context of Monte Carlo simulations where many convolutions with the same $\lmax$ and $\mmax$ are needed, it is computationally profitable to skip unneeded calculations.\footnote{Note that it is generally not more efficient to evaluate all of the $\dconv$ before hand because of the disk space required and the large amount of time needed to read them in; it is more efficient to calculate them on the fly.}  Some terms in the sum of \Eq{convdef} need not be included because their $\dconv$ are vanishingly small.  To determine which terms to exclude we turn to \Eq{classiccond} where three general possibilities are considered:
\begin{itemize}
\item $\mb$ and $\msky$ are of similar magnitude and much smaller than $l$.
\item $\mb$ and $\msky$ are of similar magnitude and of the same order of magnitude as $l$.
\item $\mb$ and $\msky$ are of widely differing magnitude with one much smaller than $l$ and one of similar magnitude.
\end{itemize}
In each of these possibilities, the neglected $\dconv$ are those evaluated at $\beta$ angles that correspond to the non-classical, exponentially suppressed region.  Before explaining how these $\dconv$ are identified, {\tt conviqt}'s nested structure should be described.  In {\tt conviqt}, the outermost loop deals with $\mb$ (which ranges from 0
to $\mmax$; nested into the $\mb$-loop
is the $\msky$-loop ranging from $-\msky$ to $\msky$; nested in the $\msky$-loop is the loop over the $\beta$ processed by that particular task.\footnote{A note to remind the reader that {\tt conviqt} is parallelized in $\beta$; each task will perform the convolutions in a subset of all the complete set of $\beta$ where $C_{\mb\msky}(\beta)$ must be calculated.}  Finally, the innermost loop is that over $l$ which is where the condition is applied.  To derive the minimum $l$ such that outside the parameter space defined by $l_{\text{min}}\le l \le \lmax$, $\dconv$ is negligible, we use \Eq{classiccond} to write
\beq \nonumber
l_{\text{min}} &=& -\frac{1}{2} + \frac{1}{2}\left[ 1 + \frac{4}{\sin^2\beta} ( \msky^2 + \mb^2 - 2 \msky\mb\cos\beta ) \right]^{1/2} \\ \label{lmin}
&\cong& \frac{\sqrt{\msky^2 + \mb^2 - 2 \msky\mb\cos\beta}- \text{\tt offset}}{\sin\beta} 
\eeq
where {\tt offset}$>$0 and ensures that the $\dconv$ neglected are well within the non-classical region and suppressed.  Calling $m_{\text{big}}$ the larger of $|\mb|$ or $|\msky|$, it is noted that $l \ge m_{\text{big}}$. To determine the {\tt offset}, we go back to the three possibilities listed above; of those, $l_{\text{min}} $ will generally equal $m_{\text{big}} $ in the cases where $\mb$ and $\msky$ are of similar magnitude; only in the third case will we generally have $l_{\text{min}} > m_{\text{big}}$, namely when $\mb$ and $\msky$ are of widely differing magnitude.  From ref.~\cite{rowe}, this case can approximately be written as a harmonic oscillator wave function:
\beq\label{asymp}
\dconv &\to& (-1)^{l-\msky}(\sqrt{l}\sin\beta_m)^{-1/2} u_{l-\msky}(\sqrt{l}(\beta-\beta_m) \\
u_\nu(x) &=& (\sqrt{\pi}2^\nu\nu!)^{-1/2}H_\nu(x)e^{-x^2/2}
\eeq
where $H_\nu(x)$ is a Hermite polynomial and $\cos\beta_m=\mb/l$.  Since we are dealing with orders of magnitude, it is not necessary to evaluate \Eq{asymp} exactly to determine {\tt offset}, only to calculate an estimate from the factor $\exp(-x^2/2)$.  We have found that using {\tt offset}$\ge\lmax/20$ gives extremely accurate results.  Finally it is noted that $d^l_{\mb\msky}(0) = \delta_{\mb\msky}$ and for that special case we put $l_{\text{min}}>\lmax$ when $\mb\ne\msky$ and no sum over $l$ is performed.
Thus, \Eq{lmin} is used to estimate whether the absolute values of all
$\dconv$ for a given combination of $l$, $\mb$, $\msky$ and $\beta$
lie below a certain threshold; if this is the case, the generation of these
values can be skipped entirely.  In particular, the most efficient version of the code written was one where the $d^{l_{\text{min}}}_{\mb\msky}$ and $d^{l_{\text{min}}+1}_{\mb\msky}$ were pre-calculated for $0\le \mb \le \mmax$, $-\lmax \le \msky \le \lmax$ and $\beta$ subset for a particular task, and read-in as seeds to the recursion relation of \Eq{lrec}.  That way, none of the unneeded $\dconv$ were calculated during the convolution.  This required an extra code to pre-compute the $\dconv$.  In the end, we opted for a single code based on the evaluation of the reduced Wigner matrix elements as described in section~\ref{wigl} because of the low overhead and maintenance as well as the high efficiency.

In this approach, we calculate all the $\dconv$ on the fly, but only include the relevant $\dconv$ in the final $l$-loop.  This code is self-contained and easier to maintain at a very minimal cost in performance.  As the $\dconv$ recursion is performed, the code checks the absolute values
of the generated $d^l_{mm'}(\beta)$ and records the $l$ index at which
a predefined threshold $\varepsilon$ (typically set to $10^{-30}$) is crossed
for the first time.
Due to the limited dynamic range of IEEE data types, values below
this threshold have no measurable influence on the convolution result and
can therefore be neglected during the final summation loop, which saves
a significant amount of CPU time.

\subsubsection{Precomputed values}

In this single code approach, a precomputation strategy well-suited to the loop structure was adopted:

\begin{itemize}
\item At the beginning of each run, we compute just once $\ln(\cos(\beta/2))$,
  $\ln(\sin(\beta/2))$ and $\cos\beta$ for all $\beta$ at
  which we need the Wigner matrix, and as mentioned above, $\ln n!$ up to
  $n=2\lmax$.
\item Also at the beginning we compute the tables
  \begin{equation*}
    P_i=\sqrt{1/(i+1)}\quad \text{and}\quad Q_i=\sqrt{i/(i+1)}
  \end{equation*}
  for $i$ in the range of 0 to $2\lmax+1$, which
  are needed for the next precomputation step.
\item inside the second loop (i.e.\ for every combination of $\mb$ and $\msky$)
  we compute the tables
  \begin{eqnarray*}
    F_{0,l}&=&(l+1)(2l+1)P_{l+\mb}P_{l-\mb}P_{l+\msky}P_{l-\msky}\text{,}\\
    F_{1,l}&=&\mb\msky/(l(l+1))\text{, and}\\
    F_{2,l}&=&Q_{l+\mb}Q_{l-\mb}Q_{l+\msky}Q_{l-\msky}(l+1)/l\text{.}
  \end{eqnarray*}
\end{itemize}

After all these preparations, eq.\ (\ref{lrec}) boils down to
\begin{equation}
  d^{l+1}_{\mb\msky}(\beta) = F_{0,l}\left(\cos\beta-F_{1,l}\right)
  d^l_{\mb\msky}(\beta) - F_{2,l}d^{l-1}_{m\msky}(\beta)\text{,}
\end{equation}
which corresponds to only five quick-to-compute floating-point operations.

The space overhead for the additional tables is $\mathcal{O}(\lmax)$,
which is insignificant compared to the $\mathcal{O}(l^2_{\text{max}})$ memory
requirement of the whole
convolution code. This also means that for the reasonable assumption of
$\lmax\lesssim 10^4$ all data required for the recursion
fit conveniently into current processors' Level-2 caches.

\subsubsection{Use of $\dlmm$ symmetries}

The use of the $\dlmm$ symmetries considerably cut the computational cost of the full sky convolution.  In particular, \Eq{pithetasym} relates the computed values of $\Cmm$ at $\beta < \pi/2$ to those at $\beta > \pi/2$.  In \Eq{pithetasym}, the phase factor $(-1)^{l}$ is accounted for by splitting the sum in \Eq{convdef} into even and odd $l$.  The phase factor $(-1)^{-m^\prime}$ is accounted for by using the relations
\beq\label{blmslmsym}
b_{lm}=(-1)^m b_{l-m}^*~, ~~s_{lm}=(-1)^m s_{l-m}^*
\eeq
and further splitting the odd and even sums of \Eq{convdef} into real and imaginary parts.  In addition, the symmetry of \Eq{mmsym} and \Eq{blmslmsym} can be used to show
\beq
C_{-\mb-\msky}(\beta) = \Cmm^*
\eeq
 speeding up the computation of $\Cmm$ by another factor of two.

\subsection{Example simulations}

To determine the accuracy of {\tt conviqt}, a detailed comparison with the stable release of the {\tt LevelS totalconvolver} \cite{Wandelt:2001gp, Reinecke:2006fv} currently compiled on the {\tt planck} cluster at the National Energy Research Science Council (NERSC) was performed.  {\tt LevelS} is a simulation package for the generation of time ordered data (TOD) by the Planck satellite~\cite{planck}.  {\tt Totalconvolver} and {\tt conviqt} both calculate a data cube that is fully compatible with LevelS.  For both codes, data cube is composed of convolved points  calculated at a polar angle $\theta$, a longitudinal angle $\phi$ and a particular beam orientation (a rotation about the beam axis) $\psi$.

\subsubsection{data cube comparison}

For $\lmax=2000$, GRASP beams for LFI-19a, $\mmax = 9$, offset = 30 and a polarized CMB map, there were 153634399 points in the data cube.  Taking the difference between the totalconvolver and conviqt data cubes (residual values below), we had:

\begin{tabular}{|c|c|c|c|c|}
\hline
{\tt offset} & avr &$\sigma$ & Max & rel \\ \hline
2000 (exact) & -1.7e-16 & 4.1e-13 & 1.4e-11 & 4.0e-8 \\ \hline
30 & -1.2e-16 & 1.3e-11 & 5.6e-10 & 1.3e-6 \\ \hline
15 & -1.6e-16 & 3.0e-9 & 1.1e-7 & 2.9e-4\\
\hline
\end{tabular}
where 'avr' refers to the average difference of the two data cubes ({\tt conviqt}($\theta,\phi,\psi$)-{\tt totalconvolver}($\theta,\phi,\psi$)), $\sigma$ is the variance of that residual data cube, 'Max' refers to the maximum value found in the residual data cube, and 'rel' refers to the ratio of $\sigma$ to the variance of the {\tt totalconvolver} data cube.  We see that with {\tt offset}=2000, the two data cubes agree to approximately 8 significant digits; for {\tt offset}=30,15 they agree to 6 and 4 significant digits respectively.

\subsection{Performance}

\begin{figure}
\resizebox{8.7 cm}{!}{\includegraphics*{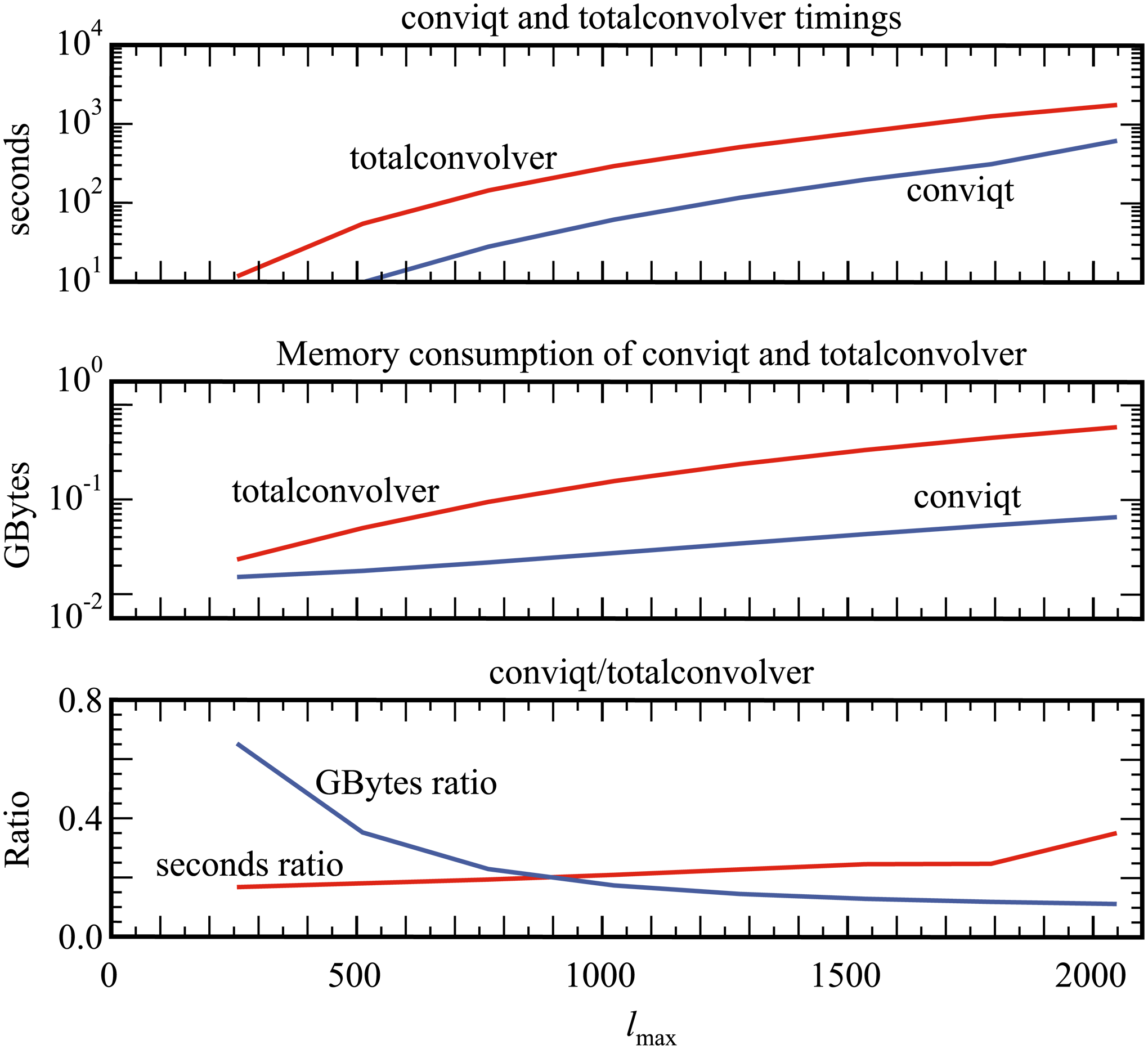}}
\caption{The top plot compares the wall clock performance of conviqt and totalconvolver at $\lmax$ intervals of 256 starting at $\lmax=256$; the middle plot compares the memory needs in GBytes of the two codes.   The lower plot shows the ratio of wall clock and Gbytes of the two codes.} \label{secgb}
\end{figure}

On a single processor on Jacquard, for lmax=2000 with a GRASP beam (LFI-19a) of $\mmax=9$, offset=30, MC=T including polarisation, conviqt had the following performance
\begin{tabular}{|c|c|c|c|c|}
\hline
{\tt Code} & Clock (secs) & gbytes \\ \hline
{\tt conviqt} & 349 & 0.37  \\ \hline
{\tt totalconvolver} & 1120  & 0.41 \\ \hline
\end{tabular}\\
where 'Clock' refers to the time the it took to complete the convolution according to the wall clock, while 'gbytes' refers to the total memory consumed.  These numbers were obtained using NERSC's Integrated Performance Monitoring (IPM) on a 712-CPU Opteron cluster called {\tt Jacquard} running a Linux operating system; each processor on {\tt Jacquard} runs at a clock speed of 2.2GHz with a theoretical peak performance of 4.4GFlop/s.  Unlike {\tt totalconvolver}, {\tt conviqt} is a massively parallel code which can be run on machines with distributed memory; running it on a single processor shows that {\tt conviqt} is intrinsically faster and more efficient than {\tt totalconvolver}.  The above table is for the case where {\tt offset}~$<$~$\lmax$, i.e., the case where {\tt conviqt} sacrifices precision for the sake of a speedier convolution.

For the general case where {\tt conviqt} and {\tt totalconvolver} calculate the same thing, we use the single code approach to obtain the following table:
\begin{tabular}{|c|c|c|c|c|}
\hline
 & \multicolumn{2}{|c|}{\tt conviqt} & \multicolumn{2}{|c|}{\tt totalconvolver} \\ \hline
 $\lmax$ & secs & GBytes & secs & GBytes \\ \hline
256 &    1.9657      &        1.51806e-02    &    11.726   &            2.32534e-02  \\ \hline
512 &    9.8691        &      1.76401e-02   &     54.634        &       5.00412e-02 \\ \hline
768 &    27.982         &     2.16856e-02        & 144.51        &      9.46579e-02 \\ \hline
1024 &  61.797         &     2.73190e-02         & 294.71     &        1.57092e-01 \\ \hline
1280 &  117.13            &  3.45383e-02      &     513.29        &    2.37350e-01 \\ \hline
1536 &  199.29           &   4.33445e-02      &     809.51        &    3.35554e-01 \\ \hline
1792 &  313.27          &    5.37376e-02     &      1264.1        &    4.51582e-01 \\ \hline
2048 &  621.26           &   6.57196e-02          & 1765.2         &   5.85376e-01 \\ \hline
\end{tabular}
These numbers are plotted in \Fig{secgb}.  The top plot shows that {\tt conviqt} is considerably faster than {\tt totalconvolver} for $\lmax<2048$; however, because both codes scale as $\lmax^3$ as $\lmax\to\infty$, the gap between their total wall clock times will narrow.  It is also seen that {\tt conviqt} consumes significantly less memory.

\begin{figure}
\resizebox{8.7 cm}{!}{\includegraphics*{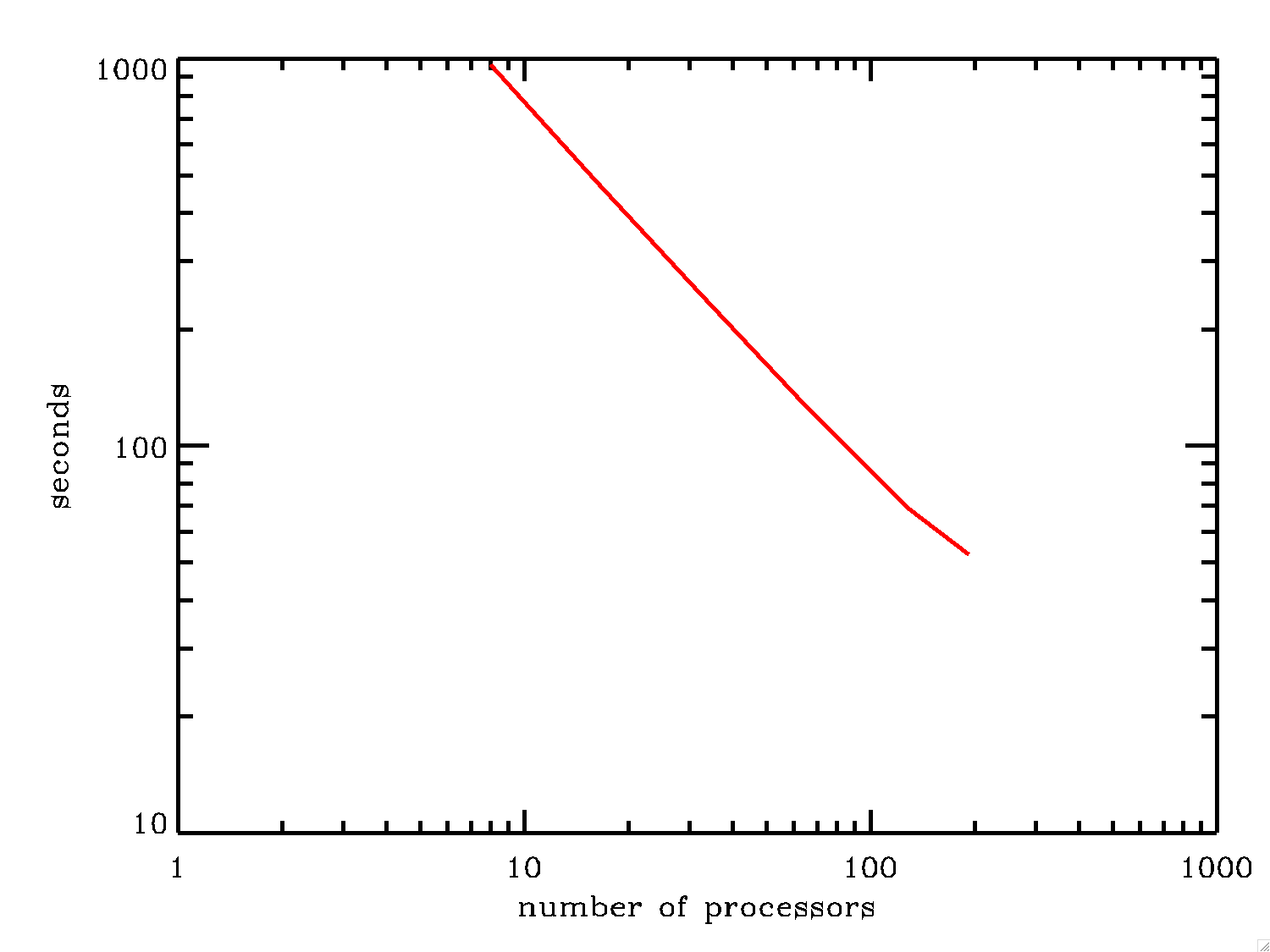}}
\caption{Timings of conviqt as the number of processors is increased with $\lmax=4096$ and $\mmax=14$.}\label{scaleNode}
\end{figure}

The scaling of {\tt conviqt} timings as a function of the total number of processors is very good.  For a $\lmax=4096$ and $\mb=14$ with polarized beam and sky, the log-log plot in \Fig{scaleNode} shows a linear relationship up to a convolution distributed on 128 processors.\\
\begin{tabular}{|c|c|}
\hline
Number of processors & seconds \\ \hline
8     &  965.46 \\ \hline
16   &  486.29 \\ \hline
32    & 247.83 \\ \hline
64    & 128.43 \\ \hline
128  & 69.06 \\ \hline
192  & 52.47 \\ \hline
\end{tabular}\\
This plot was obtained using the single code approach run on the NERSC cluster called {\tt Planck}, a 256 cores cluster of Opteron 2350 2.0GHz processors.  To measure the scaling behavior of {\tt conviqt}, no output file was created to avoid skewing the scaling law with the time it takes to write the file (tens of seconds for a 4GB file).  As the number of processors increases and the time required to perform the convolution diminishes to less than a minute, the timings become dominated with operations that have nothing to do with the convolution; among these are the reading of the input data sets (which are read in full
by all MPI tasks), the inter-process communication and various calculations which are performed
redundantly on all tasks, because communicating the results would be more expensive. Increasing the number of tasks (while keeping the problem size constant) also means a smaller number of $\beta$ angles per task,
which decreases the achievable quality of load balancing. In addition, different runs with identical inputs show variations of a few seconds in wall clock timings that have an increasing relative impact on the decreasing timings stemming from using larger numbers of processors; the most likely explanation for this are differences in the exact nature of process startup and disk access, which is not exactly reproducible in this kind of computing environment.

\section{Summary}

New algorithms for the efficient and accurate calculation of Wigner matrix elements were presented.  These algorithms were used in a full sky convolution, massively parallel algorithm called {\tt conviqt} that was shown to be significantly more efficient and much faster than the only other algorithm currently available.\\
\\
{\it \bf Acknowledgements}: GP would like to thank Maura Sandri for the GRASP 8 beams used in these simulations and Charles Lawrence for useful comments on this manuscript.  We gratefully acknowledge support by the NASA Science Mission Directorate via the US Planck Project. The research described in this paper was partially carried out at the Jet propulsion Laboratory, California Institute of Technology, under a contract with NASA.
This research used resources of the National Energy Research Scientific
Computing Center, which is supported by the Office of Science of the
U.S. Department of Energy under Contract No. DE-AC02-05CH11231.
MR is supported by the
German Aeronautics Center and Space Agency (DLR), under program 50-OP-0901, funded by the
Federal Ministry of Economics and Technology.  Copyright 2010. All rights reserved.
\begin{appendix}
\section{}\label{appa}

\beq\label{mmsym}
\dlmm &=& (-1)^{m-m^\prime}d^l_{-m-m^\prime}(\beta) \\
\dlmm &=& (-1)^{m-m^\prime}d^l_{m^\prime m}(\beta)  \\
\dlmm &=& d^l_{-m^\prime -m}(\beta)  				\\
d^l_{mm^\prime}(-\beta) &=& d^l_{m^\prime m}(\beta)  \\
d^l_{mm^\prime}(-\beta) &=& (-1)^{m-m^\prime}d^l_{mm^\prime }(\beta) \\ \label{pithetasym}
d^l_{mm^\prime}(\pi-\beta) &=& (-1)^{l-m^\prime}d^l_{-mm^\prime }(\beta) \\
d^l_{mm^\prime}(\pi-\beta) &=& (-1)^{l+m^\prime}d^l_{m-m^\prime }(\beta) 
\eeq

\end{appendix}



\end{document}